

\magnification=1200
\hoffset=0.0 true cm
\voffset=0.0 true cm
\vsize=21.5 true cm
\hsize=16.3 true cm

\baselineskip=20pt
\parskip=5pt

\parindent=22pt
\raggedbottom

\def\pp{\noindent\parshape 2 0.0 truecm 17.0 truecm 0.5 truecm 16.5 truecm}

\font\bigbf=cmb10 scaled \magstep2

\def\etal{{\it et~al.~}}
\def\lsim{\hbox{ \rlap{\raise 0.425ex\hbox{$<$}}\lower 0.65ex\hbox{$\sim$} }}
\def\gsim{\hbox{ \rlap{\raise 0.425ex\hbox{$>$}}\lower 0.65ex\hbox{$\sim$} }}
\def\pn{\par\noindent}
\def\lr{$Log R_{GC}$ }
\def\r{$R_{GC}$ }
\def\rc{$r_c$ }

\def\fd{$F_{T,D}$ }
\def\lf{$Log F_{T,D}$ }
\def\muu{$\mu_V(0)$ }
\def\mv{$|M_V|$ }
\def\ds{DSph's }
\def\d{DSph }
\def\si{$\sigma_0$ }
\null\vskip 1.0 truecm

\centerline{\bigbf ENVIRONMENTAL EFFECTS ON THE STRUCTURE}
\bigskip
\centerline{\bigbf OF THE DWARF SPHEROIDAL GALAXIES}
\bigskip
\bigskip

\centerline{\bf M. Bellazzini$^1$, F. Fusi Pecci$^2$ and F.R. Ferraro$^2$.}

\bigskip
\bigskip

\centerline{\it $^1$ Dipartimento di Astronomia, Universit\'a di Bologna,}
\centerline{\it Via Zamboni 33, I-40126 Bologna, Italy.}
\centerline{\it $^2$ Osservatorio Astronomico di Bologna, Via Zamboni 33,
I-40126 Bologna, Italy.}

\bigskip
\bigskip
\bigskip
\bigskip
\bigskip

\bigskip
\bigskip
\bigskip
\centerline{e-mail: bellazzini@astbo3.bo.astro.it}
\centerline{fax: 39-51-259407}

\bigskip
\bigskip

\centerline{To appear on {\it Monthly Notices of the Royal Astronomic Society
(Letters)}}
\bigskip

\centerline{Received: ...................................... }

\vfill\eject
\centerline{\bf ABSTRACT}
\bigskip
\noindent {The central surface brightness ($\mu_V(0)$) of the dwarf spheroidal
satellites of our own Galaxy is found to correlate with their distance from the
Galactic Center (\r). This observational evidence suggests that
environmental effects could strongly influence their structure.
This suggestion is further supported by a clear-cut bivariate correlation
linking $\mu_V(0)$ to a linear combination of $M_V$, the absolute integrated
magnitude, and $R_{GC}$, which seems to be shared also by the M31 spheroidals.
Possible implications for the Fundamental Plane of elliptical galaxies and
Dark Matter content of dwarf spheroidals are also briefly discussed.}

\bigskip\noindent

\bigskip

\centerline {\bf 1. INTRODUCTION}
\noindent
The simple fact that an unequivocal definition of the whole class of
observed dwarf spheroidal galaxies (DSph's) is still lacking may be taken
as an indication of how challenging and mysterious bodies they are.
In this note, we adopt the definitions reported by Da Costa (1992)
and Gallagher \& Wyse (1994, hereafter GW).
So the sample considered here consists of nine satellites of
our own Galaxy ({\it i.e.} Carina, Draco, Sculptor, Sextans, Fornax,
Sagittarius, Ursa Minor, Leo I, Leo II), three satellites of M31 ({\it i.e.}
And I, And II, And III) and the isolated galaxy Tucana.

As widely discussed in the exaustive reviews of Da Costa (1992), Pryor (1992),
Zinn (1993), Mateo (1993), GW, and Ferguson \& Binngeli (1994), a complete
understanding of the properties of these stellar systems is important
for many reasons, like for instance:
({\it a}) the possible relation with the formation process of their
``reference'' galaxies (Zinn 1993, van den Bergh 1994);
({\it b}) their r\^ole in the evolutionary path which leads to formation
of elliptical galaxies (Djorgovski \& de Carvalho 1990--DdC, GW);
({\it c}) their claimed high content of Dark Matter (DM) (see Pryor 1992
for a review).

Here we present a new result obtained from the analysis of the principal
correlations detectable among the observed structural parameters of
the considered sample of \ds, which adds empirical support to the hypothesis
that environmental conditions ({\it i.e.}, for instance, the tidal field of
their reference galaxy) exert a noticeable driving in settling the
intrinsic structure of \ds.

In Sect. 2 we present the data-set and the results and in Sect. 3 we
briefly discuss their implications and suggest some conclusions.

\medskip\noindent
\centerline{\bf 2. DATA-SET AND RESULTS}
\noindent
The major problem in the study of \ds
is represented by the intrinsecally low statistical significance
of any global analysis of the sample, either because
of its poorness (13 objects, at best), either because of the large
uncertainties still
associated to the measure of reliable structural parameters (radii,
integrated magnitudes, etc.) for so dim and sparse objects. Hence,
any correlation based on this sample has to be taken with particular caution.

Nevertheless, the search for correlations linking the main observables of
galaxies is a powerful tool for the study of any class of them,
and it has already been applied also to derive interesting evidences
on the \ds (DdC, GW, Mateo \etal 1993, Caldwell \etal 1993, hereafter CASD).
\pn
{\it 2.1 The data-base.}
\pn
The observables are drawn mainly from the compilation of GW (their Table 2).
The central surface brightness of Carina, Draco, And I and And II,
missing in GW, has been
taken from Mateo (1993, Table 2). The projected distances of M31 DSph
satellites are from van den Bergh (1972). The central velocity dispersions
($\sigma_0 [Km/s]$) and core radii ($r_c [pc]$) are drawn from Mateo
(1993, Table 2), with the exception of LeoII whose $\sigma_0$ value has been
superseded by a recent measure (Vogt \etal 1995).
Notice that, to get a
straightforward parameter which scales directely as the luminous mass does,
we have always used the absolute values of the integrated absolute
magnitude ($|M_V|$).
\pn
{\it 2.2 A correlation between central structure and distance from
the parent galaxy.}
\pn
Figure 1 presents the correlations occurring between \muu --the
central surface brightness in V-- and the parameters \lr --the logarithm
of the distance
from the Galactic Center-- and \mv --the integrated absolute magnitude--,
for the 8 \ds which are neighbours of the Milky Way.
At this stage, the Sagittarius \d has been excluded from the plots
to display the behaviour of an homogeneous sample of 8 objects
with no evident sign of {\it strong} tidal pertubation presently
induced by the Galaxy (but minor signs of tidal disturbance has been found also
in some other DSph; Gould \etal 1992; Gerhard 1993; Hargreaves \etal 1994).

In panel {\it (1a)}, the already studied correlation between \mv and \muu
is presented. As previously noted by GW and CASD, such a correlation is
significantly weaker than that found for the dwarf ellipticals
(see CASD, Fig. 8, but also Djorgovski \& de Carvalho 1992).

In panel {\it (1b)}, it is shown that a correlation at least as significant as
the previous one  is also standing between \lr and \muu , in the sense that
\ds closer to the Galaxy display dimmer central regions. This is the first
{\it empirical} finding, based exclusively on observed parameters, which
testifies the existence of a possible link between the intrinsic
structure of Galactic \ds and their environment.
Note that a similar trend has been shown by Mateo \etal (1991) concerning the
{\it inferred} central DM densities, assuming an isotropic velocity dispersion
and a wider distribution of the DM with respect to stars.

In order to give some assessment of the statistical significance of this
correlation we performed a numerical simulation, with conservative assumptions.
We have generated 10,000 random distributions of 8 points in the \muu-\lr
plane. For any distribution, a value of \muu ($[22-26]$) and a value of \lr
($[1.7-2.5]$)  were randomly assigned to each of the 8 points.
Then, the linear correlation coefficients between each couple of
random \muu and \lr -vectors ($r_{calc}$) were computed counting, in
particular,
how many times the absolute value of $r_{calc}$ was found to be equal or
greater than $|r_{obs}|$. Since the condition $|r_{calc}|\ge|r_{obs}|$ occurred
105 times, one can deduce that the probability that the observed configuration
in the (\muu-\lr) plane is actually drawn from a random distribution is $\simeq
1 \%$. This means that the hypothesis that \lr and \muu are actually
{\it uncorrelated} can be {\it rejected} at a level of confidence of
$\sim 99 \%$.

In panel {\it (1c)}, a bivariate correlation between \muu and a linear
combination of \mv and \lr is also shown.
As can be seen, this relation is stronger than each of the two
monovariate ones, in particular if the statistically robust Spearman
rank correlation coefficient ($s$) is considered (Press \etal 1992).
An analogous simulation as above (10,000 draws, $N[|r_{calc}|\ge|r_{obs}|]=25$)
yields a probability that \muu and the quoted linear combination of
\lr and \mv are actually uncorrelated much less than $ 1\%$.

Within this framework, it is interesting to verify the behaviour of the
Sagittarius \d (Sgr) which turns out to have a structure strongly perturbed
by the tidal field of the Milky Way (Ibata \etal 1994, Mateo \etal 1994),
and for this reason it has been excluded from the plots reported in Figure 1.

By inspecting Figure 2 (obtained with the same procedure as Figure 1
but including Sgr), one
can note that: (a) as perhaps expected,  the (\muu {\it vs.} \mv) correlation
is significantly weakened by the addition of Sgr, (b) the (\muu {\it vs.} \lr)
correlation is less affected.

On the other hand, the bivariate relation presented in panel (2c)
{\it accounts very nicely also for the position of Sgr \d in this plane}.
In other words, if one takes into account the fact that Sgr is
presently  very close to the Galaxy, Sgr does not deviate at any
significant level from the general behaviour, typical of the other
8 \ds considered in the total sample.
Adopting conservative estimates for the typical uncertainties in the
measure of the involved quantities ({\it i.e.} $\sim0.3 mag$ in \mv,
$0.3-0.5 ~mag/arcsec^2$ in \muu, and $\sim20\%$ in \r, see
GW, Mateo \etal 1993, CASD), the dispersion of this relation could
be almost completely accounted for by measurement errors, being Sextans
the only point which deviates more than one error bar from the fitting-line.

One might therefore
conclude that, in this framework, no special explanation is necessary to
account for the structure of this highly tidally disturbed \d.
So, we claim that {\it all the spheroidal satellites of the Galaxy
can share the same empirical correlation linking their structural
properties and their position with respect to the Galaxy}. Moreover,
the present claim strongly suggests that the visible central
structure of \ds is substantially controlled by their {\it luminous} mass
(or something scaling as it does) and by the Galactic tidal field.

To extend the analysis at least to a few M31 objects, we have reported as
open squares in each panel of Figure 2 the points corresponding to the
M31 satellites.  Since van den Bergh (1972) and Mateo (1993) noted
that the projected distance of AndII to M33 is significantly less than that
to M31, we also reported as a cross in (panels 2b,2c) the position
that AndII  would have if it is associated to M33 rather than to M31.
Although for these objects only the {\it projected} distance to the
reference galaxy is available, their position in the planes of panels
(2b,2c) is in fairly good agreement with the detected correlations.
This can be taken as a plausible hint that the result obtained is
not just a peculiarity of the \d satellites of our own Galaxy but
could be a quite general characteristic of all \ds orbiting around
much larger galaxies.
\pn
{\it 2.3 An important caveat: the meaning of $R_{GC}$.}
\pn
Given the quite high eccentricity inferred for the orbits of the DSph
companions to the Milky Way (GW; Oh, Lin and Aarseth 1995, hereafter OLA),
it is important to address the question of what physical parameter is
effectively measured by the present-day distance from the Galactic Center \r.
In fact, as correctly pointed out by the Referee, ``any correlation involving
the {\it present} distance would be difficult to understand as being a real
physical effect unless the present \r is an estimate of the time-averaged
distance''. In other words, the question is: why the present \r is
meaningful? Lacking reliable orbital parameters a satisfactory answer
can hardly be offered. We add below a few statistical considerations
which in our view show that, although in principle the adoption of \r could
even be misleading, in practice the probability that its use may induce
spurious correlations is very low.

First, we recall that most of the DSph's have probably completed several
orbits (at least) in one Hubble time (GW and references therein; OLA),
and that their orbital eccentricities are extimated to span the range
$[0.4;0.7]$ (OLA).

Second, we have calculated the time-averaged distance ($<R>_T$) from the center
of mass of a particle covering a Keplerian orbit with eccentricity $e=0.8$,
over the whole period $T$. Then, we have evaluated the probability that the
present randomly measured distance \r could strongly differ from $<R>_T$.

The simulation shows that $<R>_T$ cannot be overextimated by more than
$\sim 30\%$ ({\it i.e.} $R_{GC}<1.36\times <R>_T$), while the probability
that $R_{GC}$ could actually be smaller than $0.5\times<R>_T$ is $\simeq 10\%$.
With decreasing orbit eccentricity the differences become less and less
influent. Hence, most of the present-day \r are probably at least
``good indicators''
of the true $<R>_T$, and all the below considerations are based on this
assumption. Admittedly, however, the main claim of this {\it Letter}
can be confirmed only when reliable orbital parameters will be
available.
\pn
{\it 2.4 Further support to the tidal origin of the correlation.}
\pn
If the observed \muu {\it vs.} \lr correlation were settled by the Galactic
tidal field we expect that similar (or better) rank in \muu has to be
induced by some parameter directly measuring the tidal force exerted on the
central regions of \ds.

To test this hypothesis we defined a dimensionless form
of the Galactic tidal force. In a point-mass approximation (PM) the tidal
force per unit mass experienced at the core radius by a spherical satellite
in circular orbit around the Galaxy is
$$F_T={G\times {M_G\times 2r_c}\over{R_{GC}^3}}$$
where $M_G$ is the mass of the Galaxy in solar masses,
$R_{GC}$ is the distance of the
satellite from the Galactic Center,
$r_c$ is the core radius of the
satellite (both expressed in Kpc), and $G$ is the gravitational constant in the
appropriate units.
If one describes the "binding force" ($F_B$) of the satellite, per unit mass,
with:
$$F_B={G\times {M_{sat}}\over{r_c^2}}$$
where $M_{sat}$ is the total mass of the satellite,
then our dimensionless tidal force (\fd) can be defined:
$$F_{T,D}={{F_T}\over{F_B}}=\left({M_G\over M_{sat}}\right)\times
\left({r_c\over R_{GC}}\right)^3 .$$
Now, assuming a mass-to-light ratio typical of globular clusters, we take
$$M_{sat}=3\times 10^{0.4(M_{\odot V}-M_V)}$$
and after some algebra one can get an expression including just the
available observables:
$$log F_{T,D}= - 0.4|M_V| - 3log R_{GC} + 3log r_c + const.$$
Inserting then the known values for the various quantities
it is immediate to get the values of \lf for the individual
\ds included in our sample. Figure 3 (upper panel) reports a plot showing the
existence of a strong correlation between \muu and \lf
($r=-0.933; s=0.833$), closely resembling the observed ones (see Figures
1c,2c).

A very similar result (OLA) is obtained if a logarithmic potential (L) is
adopted to model a simple halo with a flat rotation curve, rather than working
in a point-mass  approximation. In this specific case the tidal force
at $r_c$ is (OLA):
$$F_T={{V_0^2\times r_c}\over{R_{GC}^2}}$$
and, with the same assumptions as above, we find
$$logF_{T,D}= -0.4|M_V| - 2log R_{GC} +3log r_c + const. .$$
\pn
Figure 3 (lower panel) shows the correlation obtained between \muu and \lf
for this second (L) case ($r=-0.933; s=0.833$).

This result cannot be conclusive either because of the poorness
of the examined sample, either because it is not demonstrated that the tidal
force of the Galaxy is strong enough to give such a relevant imprinting
on the central structure of \ds  in these remote regions.
Such a demonstration would require accurate numerical modeling, far beyond
the limits of the present analysis.

However, the N-body simulations performed by OLA seem to point toward a
conservation of the \d central density even during phases of tidal
disruption (but see also Piatek and Pryor 1995).
On this basis, they suggest that "DSph galaxies were formed with (central)
densities comparable to their present value". If this were the case, one could
imagine that the Galactic tidal field has settled the present ranking in DSph
central brightness (densities) in the early stages of their formation.
A simple exercise may add support to this claim.

Assuming a point-mass galactic potential, it is easy to calculate the
potential experienced by a test-particle located at a distance of a
core radius from the center
of a spherical galactic satellite,
under the effect of the opposite forces due to the galactic tides and
the satellite gravitation.

If one adopts for the satellite $M_S=10^7 M_{\odot}$ and core radius
$r_c=200 pc$, typical values inferred for DSphs (Mateo 1993),
and $M_G=10^{11} M_{\odot}$, one gets that the potential felt by the
particle when the satellite is located at galactocentric distance
$R_{GC}=200 Kpc$ is three times stronger than at $R_{GC}=60 Kpc$.
This implies that, keeping constant the mass of the proto-satellite,
the formation of denser \ds as well as phenomena of gas re-accretion
(Silk, Wyse and Shields 1987) could have been favoured with increasing
the galactocentric distance of the satellite by the actually stronger
``local'' gravitational field.

In summary, the above exercises simply show that the observed
characteristics of our sample {\it are compatible with} the hypothesis that
the Galactic tidal field had a primary r\^ole in settling the internal
structure of its own \d satellites.

\medskip\noindent
\centerline{\bf 3. Discussion and conclusions}
\noindent
On the basis of the results presented in this note we briefly
discuss three specific items.

\smallskip\noindent
{\it 3.1 The origin of the \muu vs. $R_{GC}$ correlation.}
\pn
The possible influence of the parent main galaxy on the evolution of its
satellites is a long-aged idea. In particular, many authors have argued that
such an influence especially holds for the \ds (see GW; OLA).
In this respect, it
is worth recalling the indication shown by van den Bergh (1994) who
noticed that the age of the bulk of the stellar populations observed in
the Galactic \ds depends on their distance from the Galactic Center.
Based on this evidence, he has also claimed that star formation in these
systems
was triggered by interaction with the Galactic corona or ``by supernova-driven
stellar wind or else by high UV flux from the young Galaxy which stripped
the gas from the closest faint satellites''.

A mechanism invoking special conditions at the early epochs of the
``Galactic system'' is probably viable also for explaining the dependence
of \d central surface brightness on \lr as here found (see also
Fergusson and Binngeli 1994, par. 1.3).

In particular Silk, Wyse and Shieds (1987) have proposed a scenario in which
the younger populations of \ds are originated from the re-accretion of
intergalactic gas. This process would be easier in regions
far from the Galactic Center, and so more distant satellites could have
exploited more important recent stellar burst resulting (perhaps) in
enhanced central brightness.
Only a mild correlation is actually present between central brightness and
relative importance of younger populations in \ds. Hence, at this stage it is
very hard to draw any firm conclusions, but the whole framework certainly
deserves future insights.

On the other hand, the fact that the strongly tidally-disturbed Sgr \d fits
nicely the empirical bivariate correlation obtained for the other
8 \ds (see Fig. 2c) leads us to explore more deeply the hypothesis
that the correlations here discussed have been mostly driven by the Galactic
tidal field. Furthermore, in sec. 2.3 we demonstrated that the observational
framework is consistent with this hypothesis, at least at a first order.

\pn
{\it 3.2 The evolutionary path of \ds and the Fundamental Plane}
\pn
Independently of the actual origin of the correlations shown in Figures
1 and 2, we can perhaps add a further consideration related to
the evolutionary path of \ds.

It is well known that \ds do not fit into the Fundamental Plane of elliptical
galaxies (DdC, Fergusson \& Binggeli 1994). In particular, ``...the same virial
theorem applies to both class of objects (\ds and normal ellipticals),
yet their scaling laws and manifolds are profoundly different.'' (DdC).
A similar statement is reported also by CASD concerning the break of
the (\muu {\it vs.} \mv) relation for dwarf ellipticals approximately
occurring at $\mu_V(0)\simeq 24$ and corresponding to the
range of parameters covered by \ds.

On the basis of the results presented in this note,
we suggest that the observed spreads or breaks produced by the inclusion
of the \ds in the Fundamental Plane of elliptical galaxies could be completely
induced by the effects resulting on the \ds due to interactions with the
main galaxy they are associated to. In other words, \ds could be the
very low-mass tail of the mass function of elliptical galaxies, whose intrinsic
scaling law has been deeply modified by environmental conditions.
As a matter of fact, \ds seem to fall under a critical mass-limit below
which the external ``imprinting'' induced by the association to a much
larger body  has prevailed on the intrinsic one.

Some support to this claim comes for instance from the evidence (DdC)
that the \d characteristic manifold is not two-dimensional, as that
of ellipticals, but {\it three-dimensional}. The additional dimension
would thus be related to the position (or the orbital parameters)
with respect to the associated large galaxy. A first direct check
on the reliability of such a scenario could perhaps be offered, for instance,
by very accurate measures of the structural parameters
of the Tucana \d, being (presently?) very far from any large galaxy
of the Local Group.

\smallskip\noindent
{\it 3.3 Is there really a lot of Dark Matter in \ds?}
\pn
Our results could also be seen as an indirect support to the idea
that the high velocity dispersions (\si) observed in \ds have tidal origin
and are not the signature of large $M/L$ ratios (see GW for references).
The "tidal hypothesis" is largely controversial and no conclusive argument
seems still able to confirm or reject it on a theoretical ground.
However, the most recent numerical simulations (Piatek and Pryor 1995; OLA)
show that the observed velocity dispersion cannot be
produced {\it only} by simple tidal interactions.
On the other hand,
Piatek and Pryor (1995) note that the complex scenario proposed by
Kuhn and Miller (1989), suggesting a tidal heating of \ds due to resonant
coupling between the time-dependent tidal field of the Galaxy and
the DSph collective oscillation modes, is not ruled out by their
simulations. We note that the mechanism proposed by Kuhn and Miller is
efficient
in modifying also the inner structure of a DSph.

Unfortunately, the paucity of accurate measures of \si in \ds does not
allow at this stage a fruitful testing of the "tidal hypothesis" from an
empirical point of view, expecially looking for a (\si vs. \r) correlation.
Figure 4 shows the distribution of the available measures in this plane:
the existence of an anti-correlation between central velocity dispersion and
\r can be guessed, if any, only in the inner $100 Kpc$
whilst the relatively high \si measured in the far LeoII galaxy seems
hardly explainable without invoking a significant dark matter component,
at least for {\it this} galaxy (Vogt \etal 1995).

Nevertheless, considering that Fornax is $\sim 3 mag$ brighter
than any other body in the sample, the possibility that Galactic tides offer
some significant contribution to increase the velocity dispersion of
\ds cannot be excluded on the basis of the present result.
If this were the case, Fig. 4 would suggest that the amount of
Dark Matter for the nearest \ds, which incidentally display the more
striking $M/L$ ratios (Mateo 1993), could have been overextimated.

\medskip\noindent
{\bf Acknowledgements.} The authors warmly thank Luca Ciotti and Alvio
Renzini for the useful discussions. They are also indebted to the referee,
Rosemary Wyse, for the helpful comments.

\bigskip\noindent
\vfill\eject

\centerline{\bf References.}
\pn
\pp Armandroff, T.E., 1993, in Proceedings of the ESO/OHP Workshop on Dwarf
Galaxies, G. Meylan and G. Prugniel eds., p. 211

\pp Caldwell, N., Armandroff, T.E., Seitzer, P., Da Costa, G.S., 1992, AJ,
103, 840 (CASD)

\pp Da Costa, G.S.,1992, in IAU Symp. n.149 "The Stellar Populations of
Galaxies", B. Barbuy and A. Renzini eds., p. 191

\pp Djorgovski, S.G., de Carvalho, R., 1990, in Windows on Galaxies, G.
Fabbiano
\etal eds., p. 9 (DdC)

\pp Djorgovski, S.G., de Carvalho, R., 1992, in Morphological and Physical
Classification of Galaxies, G. Longo \etal eds., p. 379

\pp Fergusson, H.C., Binngeli B., 1994, AAR, 6, 67

\pp Gallagher, J.S. III, Wyse, R.F.G, 1994, PASP, 106, 1225 (GW)

\pp Gerhard, 0.E., 1993, in Proceedings of the ESO/OHP Workshop on Dwarf
Galaxies, G. Meylan and G. Prugniel eds., p. 335

\pp Gould, A., Guhathakurta P., Richstone, D., Flynn, C., 1992, ApJ, 388, 345

\pp Hargreaves, J.C., Gilmore, G., Irwin, M.J., Carter, D., 1994, MNRAS,
271, 693

\pp Kuhn, J.R., 1993, ApJ, 409, L13

\pp Kuhn, J.R., Miller, R.H., 1989, ApJ, 341, L41

\pp Ibata, R., Gilmore, G., Irwin, M., 1994, Nature, 370, 194

\pp Mateo, M., 1993, in Proceedings of the ESO/OHP Workshop on Dwarf
Galaxies, G. Meylan and G. Prugniel eds., p. 309

\pp Mateo, M., Olzewski, E., Welch, D.L., Fischer, P., Kunkel, W., 1991,
AJ, 102, 914

\pp Mateo, M., Olzewski, E.W., Pryor, C., Welch, D.L., Fischer, P., 1993,
AJ, 105, 510

\pp Mateo, M., Udalski, A., Szymansky, M., Kaluzny, J., Kubiak, M., Krzeminski,
W.,1995, AJ, 109, 588

\pp Oh, K.S., Lin, D.N.C, Aarseth, S.J., 1995, ApJ, 442, 142 (OLA)

\pp Piatek S., Pryor, C., 1995, AJ, 109, 1071

\pp Press, W.H., Teukolsky, S., Vetterling, W.T., Flannery, B.P., 1992,
Numerical Recipes, 2nd. ed., Cambridge University Press, New York, USA, p. 633

\pp Pryor, C., 1992, in Morphological and Physical
Classification of Galaxies, G. Longo \etal eds., p. 163

\pp Silk, J., Wyse, R.F.G., Shields, G.A., 1987, ApJ, 322, L59

\pp van den Bergh, S., 1972, ApJ, 171, L31

\pp van den Bergh, S., 1994, ApJ, 428, 617

\pp Vogt, S.S., Mateo, M., Olzewski, E.W., Keane, M.J., 1995, AJ, 109, 151

\pp Zinn, R., 1993, in The Globular Cluster-Galaxy Connection, ASPCS vol. 48,
H. Smith and J.P. Brodie, p. 38

\vfill\eject

\centerline{\bf Figure Captions.}

\medskip\noindent
{\bf Fig. 1.} Correlations for the Galactic \ds, excluding the
Sagittarius DSph. In the plots, {\it r}
is the linear correlation coefficient, {\it s} is the Spearman's rank
correlation coefficient. The central surface brightness (\muu) is plotted
versus \mv ({\it panel 1a}) and versus \r ({\it panel 1b}), respectively.
In {\it panel 1c}, the bivariate correlation found for the quoted parameters
is shown (see Sect. 2.2).
The line represents the linear regression fit to the data, taking into
account the errors in both axes.  Error bars are based on the conservative
assumptions: $\pm 0.3~mag$ in \mv, $\pm 0.4~mag/arcsec^2$ in \muu
and $20 \%$ in $R_{GC}$, and propagating the errors when necessary.

\medskip\noindent
{\bf Fig. 2.} {\it Panels 2a, 2b, 2c}: the same as Fig. 1 but including
the Sagittarius DSph both in the plots and in the computation of
correlation coefficients. The M31 \ds are also reported (empty squares)
(see Sect. 2.2). For these galaxies, the {\it projected} distance
to M31 is actually reported.
The {\it cross} represents the location in the diagram of AndII
if it would be associated to M33 rather than to M31.

\medskip\noindent
{\bf Fig. 3.} Correlation between dimensionless tidal force
(${F_{T,D}^{'}}=F_{T,D}$ minus a constant term)
{\it vs.} central surface brightness (\muu)  for the 8 Galactic \ds
whose core radius (\rc) measures are available. The cases of point mass
potential (PM, upper panel) and logarithmic potential (L, lower panel)
are considered.
In both models the central surface brightness dims as the tidal
force increases.

\medskip\noindent
{\bf Fig. 4} Central velocity dispersions ($\sigma_0 [Km/s]$) {\it vs.} the
logarithm of the distance from the Galactic Center (\r) for the 7
\ds for which measures of $\sigma_0$ are available.
\bye